\documentclass[10pt]{iopart}
\usepackage{iopams,color,fullpage}
\usepackage[colorlinks=true,linkcolor=blue,citecolor=blue]{hyperref}
\usepackage{times,fourier,charter,graphicx}
\usepackage{amssymb}

\advance\textwidth -4mm
\advance\hoffset 2mm
\advance\textheight 2mm
\advance\voffset -1mm

\makeatletter

\eqnobysec
\def\eqref#1{(\ref{#1})}

\def\numparts{\refstepcounter{equation}%
     \setcounter{eqnval}{\value{equation}}%
     \setcounter{equation}{0}%
     \def\theequation{\arabic{section}.\arabic{eqnval}{\it\alph{equation}}}}
\def\endnumparts{\def\theequation{\arabic{section}.\arabic{equation}}%
     \setcounter{equation}{\value{eqnval}}}
\def\bse{\numparts}
\def\ese{\endnumparts}
\def\bea{\begin{eqnarray}}
\def\eea{\end{eqnarray}}
\def\be{\begin{equation}}
\def\ee{\end{equation}}

\renewcommand\section{\@startsection {section}{1}{\z@}%
  {-3.2ex \@plus -1ex \@minus -.2ex}%
  {2ex \@plus.2ex}%
  {\color{darkblue}\reset@font\normalsize\bfseries\raggedright}}
\renewcommand\subsection{\@startsection{subsection}{2}{\z@}%
  {-3ex\@plus -1ex \@minus -.2ex}%
  {1ex \@plus .2ex}%
  {\color{darkblue}\reset@font\normalsize\itshape\raggedright}}

\let\tru@int=\int
\def\int{\mathop{\textstyle\tru@int}\limits}
\def\overl@ss#1#2{\vcenter{\offinterlineskip
        \ialign{$\m@th#1\hfil##\hfil$\crcr#2\crcr<\crcr } }}
\def\overgr@at#1#2{\vcenter{\offinterlineskip
        \ialign{$\m@th#1\hfil##\hfil$\crcr#2\crcr>\crcr } }}
\def\gl{\mathrel{\mathpalette\overl@ss>}}
\def\lg{\mathrel{\mathpalette\overgr@at<}}
\def\d{\mathrm{d}}

\def\Real{\mathbb{R}}

\def\sn{\mathop{\rm sn}\nolimits}
\def\cn{\mathop{\rm cn}\nolimits}
\def\dn{\mathop{\rm dn}\nolimits}

\def\pvint{\int\kern-0.94em-\kern0.2em}
\def\sech{\mathop{\rm sech}\nolimits}

\def\dn{\mathrm{dn}}
\let\^=\hat
\let\==\bar
\let\nina=\~
\let\~=\tilde
\let\@=\mathbf
\def\_#1{\~{\@{#1}}}

\let\le=\leqslant

\def\E{E_m}
\def\K{K_m}

\def\d{\mathrm{d}}
\def\e{\mathrm{e}}

\def\txtfrac#1#2{{\textstyle\frac{#1}{#2}}}

\def\diag{\mathop{\rm diag}\nolimits}
\def\fbf#1{\setbox0=\hbox{$#1$}\kern-0.10\wd0
  \lower0.02em\copy0\kern-\wd0 \lower0.02em\hbox{\kern+0.05em\copy0}\kern-\wd0
  \raise0.02em\copy0\kern-\wd0 \raise0.02em\hbox{\kern-0.05em\box0}}
\def\bfnabla{\fbf\nabla}
\def\half{{\textstyle\frac12}}
\def\O#1{^{(#1)}}
\def\Dy[#1]{D_y{#1}}
\makeatother

\def\cn{\mathop{\rm cn}\nolimits}
\def\sn{\mathop{\rm sn}\nolimits}

\definecolor{darkblue}{rgb}{0,0,0.8}

\def\em{\endgroup}

\begin{document}
\title{\color{red}Whitham modulation theory for the Zakharov-Kuznetsov equation
and transverse instability of its periodic traveling wave solutions}
\author{Gino Biondini and Alexander Chernyavsky}
\address{Department of Mathematics, State University of New York at Buffalo, Buffalo, NY 14260}
\date{\today}
\def\submitto#1{\relax}

\begin{abstract}
We derive the Whitham modulation equations for the Zakharov-Kuznetsov equation 
via a multiple scales expansion and averaging two conservation laws 
over one oscillation period of its periodic traveling wave solutions.  
We then use the Whitham modulation equations to study the transverse stability of the periodic traveling wave solutions.
We find that all such solutions are linearly unstable, and we obtain an explicit expression for the growth rate 
of the most unstable wave numbers.
We validate these predictions by linearizing the equation around its periodic solutions
and solving the resulting eigenvalue problem numerically.
Finally, we calculate the growth rate of the solitary waves analytically.
The predictions of Whitham modulation theory are in excellent agreement with both of these approaches.
\par\medskip\noindent\today
\par\medskip\noindent\it
Dedicated to Thanasis Fokas on the occasion of his seventieth birthday.
\end{abstract}
\thispagestyle{plain}

\section{Introduction}

One of the most striking effects that can arise from the combination of dispersion and nonlinearity is the formation of 
dispersive shock waves (DSW), which are coherent, non-stationary oscillatory structures
which typically arise in the context of small dispersion problems,
and which provide a dispersive counterpart to classical shock waves \cite{Smoller1994}
(e.g., see the review~\cite{ElHoefer} and references therein).
Dispersive shock waves are known to form in surface water waves (where they are known as undular bores),
internal waves, nonlinear optics, the atmosphere, Bose-Einstein condensates, and beyond. 
Because of their ubiquity in nature, the study of DSWs continues to attract considerable interest worldwide.

A powerful tool to study small dispersion problems is Whitham modulation theory \cite{Whitham1965,Whitham1974}
(or Whitham theory for brevity). 
Looking at a DSW as a slow modulation of the periodic traveling wave solutions of the underlying 
partial differential equation (PDE), 
Whitham theory allows one to derive the so-called Whitham modulation equations (or Whitham equations for brevity), 
that govern the evolution of these periodic traveling wave solutions over longer spatial and temporal scales. 
The Whitham equations are a system of first-order, quasi-linear PDEs. 
For integrable equations in one spatial dimension, the inverse scattering transform (IST) \cite{AC1991,AS1981,NMPZ1984}
can also be used to study small dispersion limits (e.g., see \cite{LaxLevermore,BertolaTovbis,BiondiniMantzavinos,BoutetLenellsShepelsky} and references therein). 
However, Whitham theory is more broadly applicable compared to IST, 
because the former does not require integrability of the original PDE, 
and therefore it can also be applied to non-integrable PDEs. 
Moreover, even if original PDE is integrable, in many cases Whitham theory is still useful because 
it allows one to obtain a leading-order approximation of the solutions more easily.
Because of this, Whitham theory has been applied with great success to many nonlinear wave equations 
in one spatial dimension (again, see \cite{ElHoefer} and references therein). 
Until recently, however, small dispersion limits in more than one spatial dimension had been much less studied. 

Recently, one of us
derived the Whitham modulation equations for the Kadomtsev-Petviashvili (KP) equation \cite{Biondini_RSPA2017},
the Benjamin-Ono equation \cite{Biondini_PRE2017} 
and a class of equations of KP type~\cite{ABR2018}.
He then studied the properties of the resulting system of equations \cite{Biondini_NLTY2020,arXiv2303.06436}
and used it to study a variety of initial value problems
of physical interest \cite{NLTY21,PRSA22,JFM21}.
The Whitham modulation equations for the nonlinear Schr\"odinger (NLS) equation 
in two \cite{ACR2023} and three \cite{Abeya2023} spatial dimensions were also recently derived.
In this work we continue this program of study, aimed at generalizing and applying Whitham modulation theory
to nonlinear wave equations in two and three spatial dimensions.
Specifically, we derive the Whitham modulation equations for another physically relevant model, 
namely, the Zakharov-Kuznetsov equation,
and we use the resulting system of equations to study the transverse stability of its periodic traveling wave solutions.

The Zakharov-Kuznetsov (ZK) equation \cite{ZK} is a physical model arising in many different contexts,
including fusion plasmas and geophysical fluids \cite{InfeldRowlands},
magnetized plasmas \cite{LannesLinaresSaut,ZK}, vortex soliton theory \cite{Nozaki}
and wave turbulence \cite{Nazarenko}.
In $N$ spatial dimensions and in the semiclassical scaling, the ZK equation is written as
\vspace*{-0.6ex}
\be
u_t + u u_{x_1} + \epsilon^2(\Delta u)_{x_1} = 0,
\label{e:ZK}
\ee
where $\@x = (x_1,\dots,x_N)$ are the spatial coordinates,
$\Delta = \partial_{x_1x_1} + \cdots + \partial_{x_Nx_N}$ is the Laplacian operator,
and $0<\epsilon\ll 1$ is a small parameter that quantifies the relative magnitude of
dispersive effects compared to nonlinear ones.
Note that the first spatial coordinate plays a special role compared to the other ones.
Accordingly, for brevity we will simply write $x = x_1$ below. 
When solutions are independent of $x_2,\dots,x_N$,
the ZK equation~\eqref{e:ZK} reduces to the celebrated Korteweg-de Vries (KdV) equation. 
Therefore, the ZK equation is, like the Kadomtsev-Petviashvili (KP) equation, a multi-dimensional generalization of the KdV equation.
Unlike the KdV equation and the KP equation, however,
the ZK equation does not appear to be integrable.
(To avoid confusion, we mention that \cite{Nazarenko} refers to \eqref{e:ZK} as the Petviashvili equation.)

The well-posedness of certain initial value problems and initial-boundary value problem for \eqref{e:ZK} was studied in \cite{arXiv:2306.07433,Herr2020,Kinoshita2021,Linares2010,Saut2010},
and the decay rate of localized solutions was studied in \cite{Mendez2021,Mendez2023}.
The stability of its solitary wave solutions was studied with various methods 
by several authors \cite{BettinsonRowlands,Bridges2000,Cote2016,arXiv:2006.00193,Klein2021,Klein2023,Pelinovsky2018,Yamazaki2017},
and that of its periodic solutions was studied in \cite{Johnson2010}.
Finally, a wave kinetic equation for \eqref{e:ZK} was derived using formal methods in \cite{Nazarenko}
and rigorously in \cite{arxiv2106.09819} for a stochastic perturbation of \eqref{e:ZK} on a lattice.

Despite its similarities with the KP equation, the ZK equation~\eqref{e:ZK} is not of KP type in the sense of \cite{ABR2018},
because~\eqref{e:ZK} is fully evolutionary, i.e., no auxiliary field is present.
Therefore the methodology presented in \cite{ABR2018} does not apply.
Specifically, the ZK equation~\eqref{e:ZK} differs from the KP equation in two important respects: 
(i) the terms involving derivatives with respect to the transverse variables $x_2,\dots,x_N$
contain third-order derivative, not second-order ones, and 
(ii) these terms involve mixed derivatives.
We will see that, as a result, 
the parametrization of the traveling wave solutions of the ZK equation is quite different from that of 
the solutions of the KP equation, 
and in fact it has some similarities with the periodic solutions of two-dimensional NLS equation.
Indeed we will see that the Whitham modulation system for the ZK~equation contains a mix of the features
of the systems for the KP and NLS equations.

The main result of this work is the ZK-Whitham system (ZKWS) of modulation equations~\eqref{e:ZKWS},
or equivalently~\eqref{e:ZKWScomp}, 
as well as a transverse stability analysis of the periodic traveling wave solutions of the ZK equation~\eqref{e:ZK}.
In section~\ref{s:derivation} we present the derivation of the ZKWS.  
In particular, in \ref{s:periodic} we introduce the periodic traveling wave solutions and relevant conservation laws of \eqref{e:ZK},
in \ref{s:multiplescales} we present the multiple scales expansion used for the derivation, 
in \ref{s:averages} we present the relevant period averages,
in \ref{s:ZKWS} we present the calculations needed to obtain the ZKWS in final form,
and in \ref{s:symmetries} we discuss some basic symmetries and reductions of the ZKWS.
In section~\ref{s:stability} we then use the ZKWS to study the transverse stability of the periodic traveling wave solutions,
and we validate the predictions of Whitham theory by comparing them with two alternative approaches.
Section~\ref{s:conclusions} offers some concluding remarks.

\section{Whitham modulation theory for the ZK equation}
\label{s:derivation}

\subsection{Periodic traveling wave solutions and conservation laws of the ZK equation}
\label{s:periodic}

Recall that the Whitham equations describe modulations of periodic solutions of a nonlinear PDE.
Therefore, the first step in formulating Whitham modulation theory is to write down the 
periodic solutions of the PDE.
The ZK equation~\eqref{e:ZK} admits periodic traveling wave solutions, which 
are most conveniently expressed by introducing Riemann-type variables $r_1 \le r_2 \le r_3$,
similarly to what is done for the KdV, KP and NLS equations.
The derivation of these solutions is similar to that for the periodic solutions of those equations,
so we omit the details for brevity.
However, one can easily verify that \eqref{e:ZK} admits the following ``cnoidal wave'' solutions:
\be
u(\@x,t) = (1+q^2)\left[ (r_1-r_2+r_3) + 2(r_2-r_1)\cn^2(2K_m Z,m) \right],
\label{e:periodicsolution}
\ee
where $\cn(z,m)$ is the Jacobi elliptic cosine \cite{NIST}, 
$K_m = K(m)$ the complete elliptic integral of the first kind,
\bse
\be
m = \frac{r_2-r_1}{r_3-r_1}
\label{e:mdef}
\quad
\ee
is the elliptic parameter, 
\bea
Z=(\@k\cdot\@x - \omega t)/\epsilon\,,\qquad
\@k = (k_1,\dots,k_N)\,,\qquad \@q = (k_2,\dots,k_N)/k_1\,,
\\
k_1 = \frac{\sqrt{r_3 - r_1}}{2\sqrt6 K_m}, 
\quad
\omega = \txtfrac 13(1+q^2)(r_1 + r_2 + r_3)\,k_1\,,
\label{e:klomega}
\eea
\ese
and $q^2 = \@q\cdot\@q = q_1^2 + \cdots + q_{N-1}^2$.
The solution~\eqref{e:periodicsolution} is uniquely determined by $N+2$ independent parameters, $r_1,\dots,r_3$ and $q_1,\dots,q_{N-1}$,
and it describes wave fronts localized along the lines
$\@k\cdot\@x - \omega t = 2n\pi$,
with unit period with respect to the variable $Z$
and period $2K_m$ with respect to the variable $x$.
Note the appearance of the factor $1+q^2$ in \eqref{e:periodicsolution} and \eqref{e:klomega},
unlike the KP equation \cite{Biondini_RSPA2017},
and similarly to the NLS equation in $N$ spatial dimensions~\cite{Abeya2023}.
In keeping with the notation for the first spatial coordinate, we will simply write $k_1 = k$.
Also, when there are only two spatial dimensions (i.e., $N=2$), we will simply write 
$y = x_2$, $l = k_2$ and $q = q_1$.

The above solutions admit two nontrivial limits: the harmonic limit, obtained when $m=0$, corresponding to $r_2=r_1$,
and the soliton limit, obtained when $m=1$, corresponding to $r_2=r_3$.
Specifically, recalling that $\cn(z,m) = \cos z + O(m)$ as $m\to0$ and $\cn(z,m) = \sech z + O(1-m)$ as $m\to1^-$,
it is trivial to see that, as $m\to0$, the solution~\eqref{e:periodicsolution} describes vanishing-amplitude harmonic 
oscillations on a non-zero background, whereas, as $m\to1$, the solution limits to the line soliton solutions of the ZK~equation.
Explicitly, in two spatial dimensions,
\be
u_s(\@x,t) = (1+q^2)\big[ \=u + 12c\sech^2\big(\sqrt{c}(x + qy - Vt\big) \big],
\label{e:solitonsolution}
\ee
where $\=u = r_1$, $c = (r_3 - r_1)/6$ and $V = (1+q^2)\,(\=u + 4c)$. 
However, we emphasize that the modulation theory presented below applies to all of the periodic solutions~\eqref{e:periodicsolution}.

Recall that several methods can be used to derive the Whitham equations: multiple scales perturbation theory 
(as in \cite{Biondini_RSPA2017}), 
averaged Lagrangians \cite{Whitham1965}, 
and averaged conservation laws (as in \cite{Abeya2023}). 
Here we will employ the latter.
Accordingly, we need the conservation laws of the ZK equation~\eqref{e:ZK}.
The ZK equation itself can be written as a conservation law in differential form:
\bse
\be
u_t + \big( \half u^2 + \epsilon^2\Delta u \big)_x = 0\,.
\ee
Note that in this case there is no flux along the coordinates $x_2,\dots,x_N$.
Moreover, the ZK equation admits an additional differential conservation law
related to conservation of mass:
\be
(u^2)_t + \big[\txtfrac23 u^3 + 2\epsilon^2 \big(u\Delta u - u_x^2  + (\bfnabla_\perp u)^2\big)\big]_x 
  - 2\epsilon^2\,\,\bfnabla_\perp\cdot(u_x\bfnabla_\perp u) = 0\,,
\ee
\ese
where $\bfnabla_\perp = (\partial_{x_2},\dots,\partial_{x_N})$
is the gradient with respect to the transverse variables.
As mentioned earlier, the ZK equation is not completely integrable, 
unlike the KdV and KP equations, so only a limited number of conservation laws are available.
Nonetheless, below we will show that the above conservation laws will be sufficient for the derivation of 
the Whitham modulation equations.

\subsection{Multiple scales expansion}
\label{s:multiplescales}

As usual in Whitham theory, we now look for modulations of the above periodic solutions.
Specifically, we introduce the fast variable $Z$ defined by 
\be
\bfnabla_{\@X} Z = \frac{\@k}\epsilon,\qquad
Z_t = - \frac\omega\epsilon,
\label{e:Zfast}
\ee
as well as the slow variables $\@X = \@x$ and $T = t$, 
and we look for solutions 
\be
u(\@x,t) = u(Z,\@X,T), 
\label{e:ansatz}
\ee
where all of the solution's parameters are now functions of $\@X = (X_1,\dots,X_N)$ and $T$.
In particular, $\@k$ and $\omega$ are now the local wavevector and the local frequency.
Recall that in two spatial dimensions we have 4 independent parameters: 
$r_1,r_2,r_3$ and $q = q_1$.
With the above multiple scales ansatz, one has
\be
\bfnabla_{\@x} = \frac{\@k}\epsilon \partial_Z + \bfnabla_{\@X},\qquad
\partial_t = - \frac\omega\epsilon \partial_Z + \partial_T\,.
\ee
Or, in two spatial dimensions, simply
$\partial_x = (k/\epsilon)\partial_Z + \partial_X$,
$\partial_y = (l/\epsilon)\partial_Z + \partial_Y$ and
$\partial_t =  - (\omega/\epsilon)\partial_Z + \partial_T$,
with $X = X_1$ and $Y = X_2$.
Inserting the above ansatz into \eqref{e:ZK}, to leading order one recovers the periodic solutions in section~\ref{s:periodic},
but where the parameters $r_1,r_2,r_3$ and $\@q$ are now functions of $\@X$ and $T$.
The Whitham modulation equations that we are seeking are precisely the PDEs that govern 
the spatio-temporal dynamics of these solution parameters.

It is clear from the above discussion that one needs $N+2$ equations to obtain a closed system.
The first few Whitham modulation equations, referred to as ``conservation of waves'',
are simply a consequence of the above ansatz and cross-differentiability of $Z$:
\bse
\label{e:conservationofwaves}
\bea
\@k_T + \bfnabla_{\@X}\omega = 0\,,
\label{e:dkvecdt}
\\
\bfnabla_{\@X}\wedge\@k = 0\,,
\label{e:constraint}
\eea
\ese
where $\@v\wedge\@w$ denotes the $N$-dimensional wedge product, which in two and three spatial dimensions can be replaced by the standard cross product \cite{Frankel}.
In two spatial dimensions, recalling that $l = qk$, \eqref{e:dkvecdt} becomes
\bse
\bea
k_T + \omega_X = 0,
\label{e:dkdt}
\\
(kq)_T + \omega_Y = 0\,,
\label{e:dqdt}
\\
\noalign{\noindent while \eqref{e:constraint} becomes}
k_Y = (kq)_X\,.
\label{e:k_y=l_x}
\eea
\ese
Equation~\eqref{e:dkvecdt} above provides $N$ evolution equations, 
whereas, similarly to \cite{Biondini_RSPA2017,Abeya2023},
\eqref{e:constraint} provides constraints on the initial values of the dependent variables
(whose role will be discussed more fully below).
Since 
we need $N+2$ modulation equations, 
one must therefore supplement~\eqref{e:dkvecdt} 
by obtaining two additional modulation equations.
The simplest way to do that is to average the first and second conservation laws
over one spatial period,
obtaining
\bse
\label{e:modulations}
\bea
\overline{u_T^{\phantom a}\kern0em} + \overline{u^{\phantom a}\kern-0.4em u_X} + \epsilon^2\overline{\Delta u_X} = 0, 
\\[0.6ex]
\overline{(u^2)_T}
  + \overline{\big[\txtfrac23 u^3 + \epsilon^2\big(2 u\Delta u - u_X^2+(\bfnabla_\perp u)^2\big)]_X}
  - 2\epsilon^2\,\,\overline{\bfnabla_\perp\cdot(u_X\bfnabla_\perp u^{\phantom1}\kern-0.4em)} = 0,
\eea
\ese
where $\bfnabla_\perp = (\partial_{X_2},\dots,\partial_{X_N})$ is the transverse gradient in the slow variables,
and where throughout this work the overbar will denote the integral of a quantity with respect to $Z$
over the unit period.
The next step in the derivation of the modulation equations is therefore to compute the above period averages.

\subsection{Period averages}
\label{s:averages}

\begin{figure}[t!]
\centerline{\includegraphics[width=0.75\textwidth]{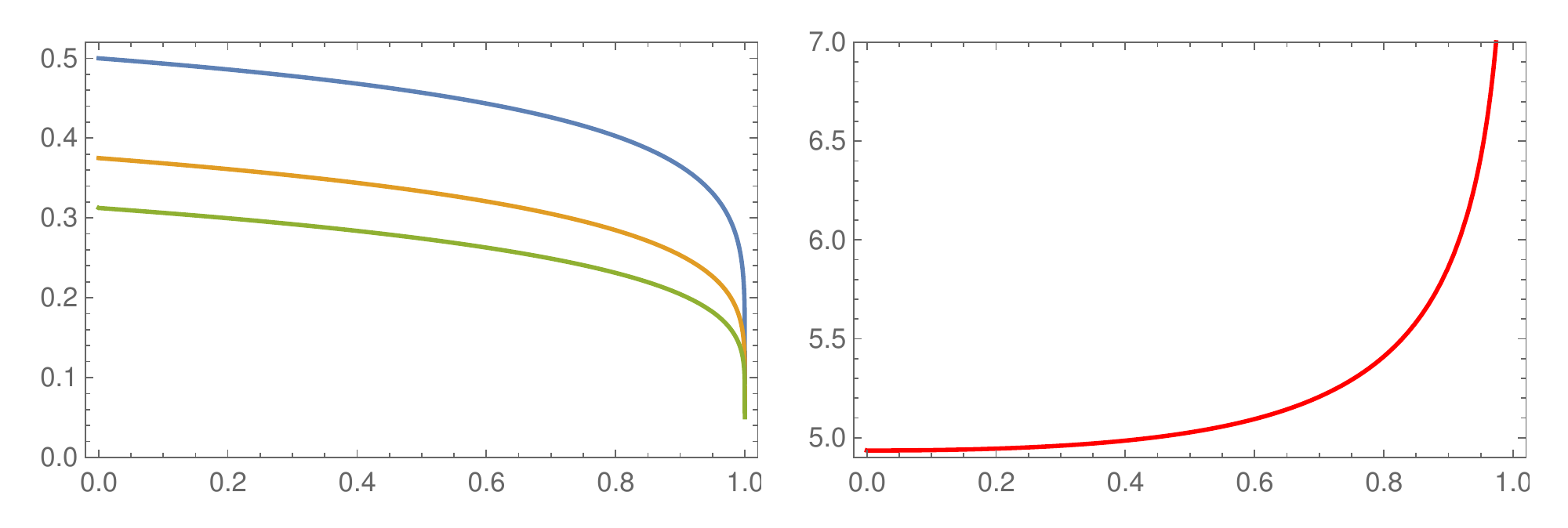}}
\caption{The quantities $G_1(m),\dots,G_4(m)$ in~\eqref{e:G1-G4} 
(in green, orange, blue and red, respectively) as a function of~$m$.}
\label{f:1}
\end{figure}

Inserting 
the ansatz~\eqref{e:ansatz},
the leading-order solution~\eqref{e:periodicsolution}
and using \eqref{e:Zfast},
to leading order the averaged conservation laws~\eqref{e:modulations} yield
\bse
\label{e:avgconslaws}
\bea
(\overline{u})_T + \left(\half \overline{u^2}\right)_X  = 0, 
\label{e:conslaw1}
\\
(\overline{u^2})_T + \left(\txtfrac23\overline{u^3} - k^2 (3+q^2)\overline{(u_Z)^2}\right)_X
  - 2\bfnabla_\perp\cdot\bigl(k^2 \@q \overline{(u_Z)^2}\bigr) = 0\,.
\label{e:conslaw2}
\eea
\ese
All of the integrals appearing in the above averages can be computed exactly, yielding~\cite{ByrdFriedman}
\bse
\label{e:uaverages}
\bea
\overline{u} = (1+q^2)\big[ r_1-r_2+r_3 + 2(r_2-r_1) G_1 \big]\,,
\\
\overline{u^2} = (1+q^2)^2\big[ ( r_1-r_2+r_3 )^2 + 4( r_1-r_2+r_3 )(r_2-r_1)G_1 + 4(r_2-r_1)^2 G_2 \big]\,,
\\
\overline{u^3} = (1+q^2)^3\big[ ( r_1-r_2+r_3 )^3 + 6( r_1-r_2+r_3 )^2(r_2-r_1)G_1
\kern14em 
\nonumber\\
\kern14em 
  + 12(r_1-r_2+r_3)(r_2-r_1)^2G_2 + (r_2-r_1)^3 G_3 \big],
\\
\overline{(u_Z)^2} = (1+q^2)^2 4(r_2-r_1)^2 G_4\,,
\eea
\ese
where 
\bse
\label{e:G1-G4}
\bea
G_1(m) = \int_0^1\cn^2(2K_m z,m)\,\d z = \frac{E_m - (1-m)K_m}{mK_m}\,,
\\
G_2(m) = \int_0^1\cn^4(2K_m z,m)\,\d z = \frac{-2(1-2m)E_m + (2-5m+3m^2)K_m}{3m^2K_m}\,,
\\
G_3(m) = \int_0^1\cn^6(2K_m z,m)\,\d z = \frac{(8-23m(1-m))E_m - (1-m)(8 - 19m +15m^2)K_m}{15m^3K_m}\,,
\nonumber\\
\\
G_4(m) = 16 K_m^2\int_0^1 \cn^2(2K_m z,m)\dn^2(2K_m z,m)\sn^2(2K_m z,m)\,\d z 
\nonumber\\
\kern4em
  = 16 K_m \frac{2(1-m(1-m))E_m - (1-m)(2-m)K_m}{15m^2}\,,
\eea
\ese
and $E_m = E(m)$ is the complete elliptic integral of the second kind.
The behavior of these quantities as a function of~$m$ is shown in Fig.~\ref{f:1}.
Their limiting values as $m\to0$ are 
\be
G_1(0) = 1/2,\quad G_2(0) = 3/8,\quad G_3(0) = 5/16,\quad G_4(0) = \pi^2/2\,,
\ee
while their asymptotic behavior as $m\to1$ is
\bse
\bea
G_1(m) = - \frac{2 + o(1)}{\log(1-m)},~
G_2(m) = - \frac{4 + o(1)}{3\,\log(1-m)},~
G_3(m) = - \frac{16+o(1)}{15\,\log(1-m)},\quad
\\
G_1'(m) = - \frac{1+o(1)}{2(1-m)K_m^2},~
G_2'(m) = - \frac{1+o(1)}{3(1-m)K_m^2},~
G_3'(m) = - \frac{4+o(1)}{15\,(1-m)K_m^2},~
\\
G_4(m) = - \frac{32+o(1)}{15K_m},\quad
G_4'(m) = - \frac{16+o(1)}{15(1-m)},
\eea
\ese
as $m\to1$.  
Also recall that $K_m = - \half(\log(1-m)-4\log2) + O(1-m)$
and $K_m' = (E_m - (1-m)K_m)/(2m(1-m)) = 1/(2(1-m)) + \frac18(\log(1-m)-4\log2+3) + O(1-m)$ 
as $m\to1$
\cite{NIST}.
These singular behaviors as $m\to1$ imply that certain rescalings are needed in order to write the modulation equations
in a convenient form, as discussed below.

\subsection{The ZK-Whitham system}
\label{s:ZKWS}

For brevity we will only write down explicitly the modulation equations in detail in two spatial dimensions,
but we emphasize that the calculations below are trivially generalized to any number of transverse dimensions, 
in a similar manner as in~\cite{Abeya2023}.
Also, 
for simplicity from now on we will write derivatives with respect to $X,Y$ and $T$ simply as derivatives with respect to $x,y$ and $t$.

Using the averages~\eqref{e:uaverages},  
recalling the definition of $k$ and $\omega$ in~\eqref{e:klomega},
and collecting all terms, 
equations \eqref{e:dkdt}, \eqref{e:dqdt}, \eqref{e:conslaw1} and \eqref{e:conslaw2}
yield a system of four modulation equations.
As~usual, however, some manipulations are needed in order to write the resulting system in the most convenient form.
We turn to this issue next.

We begin with the first conservation of waves equation, namely \eqref{e:dkdt}.
Recalling \eqref{e:klomega}, multiplying \eqref{e:dkdt} by $(1-m)K_m/k$ one then obtains an expression that remains finite
both as $m\to0$ and $m\to1$.

The second conservation of waves equation requires some additional treatment.
In this case, one can first use~\eqref{e:dkdt} to replace $k_t$, 
obtaining, as in \cite{Abeya2023,ABR2018,Biondini_RSPA2017},
the universal transversal modulation equation
\be
q_t + (D_y\omega)/k = 0\,, 
\label{e:qtnew}
\ee
where 
\be
D_y = \partial_y - q \partial_x\,,
\label{e:Dy}
\ee
is the convective derivative, which will appear prominently in all modulation equations below, 
similarly to other modulation systems in two spatial dimensions \cite{Abeya2023,ABR2018,Biondini_RSPA2017}.
Unlike the first conservation of waves equation, however, in this case in order to obtain a non-trivial equation
in the limit $m\to1$ it is necessary to use the constraint \eqref{e:k_y=l_x}, 
which we can rewrite so that it remains finite as $m\to0$ and $m\to1$ as
\be
c_1\Dy[r_1] + c_2\Dy[r_2] + c_3\Dy[r_3] + c_4q_x = 0\,,
\label{e:qconstraint}
\ee
with
\be
c_1 = (1-m)(\K-\E),~ 
c_2 = \E - (1-m)\K,~ 
c_3 = - m\E,~
c_4 = 2(r_2-r_1)(1-m)\K\,.
\ee
Then, subtracting $\omega/(k\K)$ times~\eqref{e:qconstraint} from \eqref{e:qtnew} we finally obtain the desired
modulation equation.

The averaged conservation laws~\eqref{e:avgconslaws} are the most complicated,
as can be seen from \eqref{e:uaverages} and~\eqref{e:G1-G4}.
The only manipulation needed to regularize the resulting equations, however, is just multiplication by $(1-m)\K$.  
In light of~\eqref{e:uaverages}
it is also convenient to divide \eqref{e:conslaw1} by $(1+q^2)$ and \eqref{e:conslaw2} by $(1+q^2)^2$,
respectively.
(One could also subtract a linear combination of the first conservation law and the first conservation of waves 
from the second conservation law to try to simplify it, but this is unnecessary for the present purposes.)

The collection of the resulting four modulation equations can be written in matrix form as
\be
\~C\,\@r_t + \~A\,\@r_x + \~B\,\@r_y = \@0\,,
\label{e:ZKWS0}
\ee
where $\@r = \@r(x,y,t) = (r_1,r_2,r_3,q)^T$ collects the four dependent variables.
Specifically, we write the first row of~\eqref{e:ZKWS0} from \eqref{e:dkdt}, 
the second and third row from \eqref{e:conslaw1} and \eqref{e:conslaw2}, respectively,
and the fourth row from \eqref{e:dqdt}. 
Hereafter, $\@0_{m\times n}$ and $\@1_{m\times n}$ denote matrices of size $m\times n$ with 
all entries equal to 0 or 1, respectively,
and for brevity we will drop the size notation when it should be clear from the context.

All entries of the coefficient matrices $\~C$, $\~A$ and $\~B$ in~\eqref{e:ZKWS0} 
are finite for all $0<m<1$ as well as in the limits $m\to0$ and $m\to1$.
On the other hand, their explicit expressions are fairly complicated,
and are therefore omitted for brevity, since they are just an intermediate step in the derivation.
At the same time, we next show how one can considerably simplify the system by suitably diagonalizing 
the coefficients of the temporal derivatives.

Owing to \eqref{e:qtnew},
the last row of $\~C$ is simply $(0,0,0,1)$.
Writing $\~C$ in block diagonal form, 
it is therefore convenient to 
introduce a partial inverse of $\~C$ as 
$C^{-1} = (\~C_{3\times3}^{-1},1)$,
where $\~C_{3\times3}$ denotes the upper-left $3\times3$ block of $\~C$.
Multiplying \eqref{e:ZKWS0} from the left by $C^{-1}$, 
one can then solve the above system of modulation equations for the temporal derivatives, 
which yields the final \textit{ZK-Whitham system} (ZKWS) in matrix form as 
\be
\@r_t + A\,\@r_x + B\,\@r_y  = \@0,
\label{e:ZKWS}
\ee
where the coefficient matrices
$A = C^{-1} \~A$ and $B = C^{-1} \~B$
are
\vspace*{0.6ex}
\bea
A = \left(\begin{array}{cc}
  (1+q^2)\,V_\mathrm{diag} & \frac2{45}q\,A_{3\times 1} \\
  \@0_{1\times 3} & (1+q^2)\,V 
  \end{array}\right) - q B, 
\qquad
B = \left(\begin{array}{cc}
  B_{3\times 3} & B_{3\times 1} \\
  \txtfrac13(1+q^2)\@1_{1\times3} & 2q\,V
  \end{array}\right), 
\eea
with 
\bse
\be
V_\mathrm{diag} = \diag(V_1,\dots,V_3), 
\ee
where $V_1,\dots,V_3$ are velocities of the KdV-Whitham system, namely
\vspace*{0.6ex}
\be
V_1 = V - 2b \frac{\K}{\K - \E},~
V_2 = V - 2 b \frac{(1-m) \K}{\E - (1-m) \K},~
V_3 = V  + 2b \frac{(1-m) \K}{m\E},~
\ee
\ese
where $V = \txtfrac13(r_1 + r_2 + r_3)$, $b = 2(r_2 - r_1)$ is the amplitude of oscillations in~\eqref{e:periodicsolution},
and with 
\bse
\bea
A_{3\times1} = D_{3\times3}^{-1}\@a,
\\
B_{3\times1} = \frac4{45}\frac{1+5q^2}{1+q^2}(r_3-r_1)^2\,b_o\,D_{3\times3}^{-1}\@e,
\\
B_{3\times3} = \frac{4q}{45m\K}\,(r_3-r_1)\,D_{3\times3}^{-1}\,\@e^T \!\!\otimes\@b\,,
\\
D_{3\times3} = \diag(\@d),\qquad
\eea
\ese
where 
$\@v^T\!\otimes\@w$ denotes the outer product of two vectors, namely $(\@v^T\!\otimes\@w)_{i,j} = v_iw_j$, with
\be
\@a = (a_1,a_2,a_3)^T\,,\quad \@b = (b_1,b_2,b_3)^T,\quad 
\@d = (d_1,d_2,d_3)^T,\quad 
\@e = (-1,1,1)^T\,, 
\ee
and, finally, with
\bse
\bea
a_1 = ((1 + m(14+m))\E - \K (1-m)(1+7m)\K)(r_3-r_1)^2 + 45d_1 r_1^2\,,
\\
a_2 = -((1 - m(16 + 29m))\E - (1-m)(1 - m (8 + 45m))\K)(r_3-r_1)^2 + 45 d_2 r_1 (2r_2 - r_1)\,,
\nonumber\\
\\
a_3 = (8(2-m)(1-m)\K + (29 + m(16-m))\E)(r_3-r_1)^2 + 45 d_3 r_1 (2r_3 - r_1)\,,
\\
b_o = (2-m)(1-m)\K - 2 (1-m(1-m))\E\,,
\\
b_1 = 2 (1+2 m^2)\E\K-(1-m) (1+2m)\K^2-(1-m+m^2)\,\E^2,
\\
b_2 = (1-m)(1-3m)\K^2 - 2 (1 - m (2-3 m)) \E\K+\frac{m^2-m+1}{1-m}\E^2,
\\
b_3 = m \left( 5 (1-m) \K^2 - 2 (2-m) \E\K -\left(\frac{1}{1-m} - m \right) \E^2\right),
\\
d_1 = \K - \E,~~
d_2 = \E - (1-m)\,\K,~~
d_3 = \E\,.
\eea
\ese
Equivalently, in component form, the ZKWS~\eqref{e:ZKWS} comprises the following four PDEs:
\bse
\label{e:ZKWScomp}
\vspace*{0.6ex}
\bea
r_{j,t} + (1+q^2) V_j\,r_{j,x}
  + b_4 \,D_y r_j 
  + h_j\,q_x 
  + \nu_j\, D_y q = 0,
\qquad j = 1,2,3,
\\
q_t + (1+q^2)V\,q_x + (1+q^2)\,V_x + 2qV\,\Dy[q] = 0\,,
\eea
\ese
where $\Dy[]$ is the convective derivative introduced in \eqref{e:Dy}, and
\bea
b_4 = \frac{4q (r_3-r_1)e_j}{45m K_m d_j}\,\@b\cdot\@r,\quad
h_j = \frac{2q}{45}\frac{a_j}{d_j},\quad
\nu_j = \frac{4(1+5q^2)(r_3-r_1)^2 b_o e_j}{45d_j(1+q^2)},
\quad
j=1,2,3\,.
\eea

We should point out that while the algebraic manipulations described above are rather tedious, 
they are nonetheless straightforward,
and are easily carried out with any computer algebra software.
We also point out that the ZK-Whitham system~\eqref{e:ZKWS} (ZKWS) is considerably simpler than what one would obtain
by multiplying \eqref{e:ZKWS0} by the full inverse of~$\~C$.
More importantly, note how the above ZKWS is purely in evolution form
(i.e., all four equations contains a temporal derivative), 
like those for the two- and three-dimensional NLS equations \cite{Abeya2023,ACR2023},
and unlike those for the KP equation \cite{Biondini_RSPA2017}, two-dimensional Benjamin-Ono equation \cite{Biondini_PRE2017}
and modified KP equation \cite{ABR2018}.
This is of course a direct consequence of the fact that the ZK equation~\eqref{e:ZK} does not comprise a spatial constraint 
like the KP equation and the two-dimensional Benjamin-Ono equation.

\subsection{Symmetries, reductions and distinguished limits of the ZKWS}
\label{s:symmetries}

Like the Whitham modulation systems for the KdV, KP and NLS equations, the ZK-Whitham system~\eqref{e:ZKWS} 
admits a number of symmetries and reductions.

\paragraph{Symmetries.}
The ZK-Whitham system preserves some of the physical symmetries of the ZK equation,
specifically, the symmetries under space-time translations and scaling: 
\bse
\bea
u(\mathbf{x},t) &\mapsto u(\mathbf{x}-\mathbf{x}_0,t-t_0), \\
u(\mathbf{x},t) &\mapsto a^2 u(a\mathbf{x},a^3 t), 
\eea
\ese
respectively,
where $a,t_0$ are arbitrary real constants, $\mathbf{x}_0$ is an arbitrary $N$-component real vector. 
The ZK equation~\eqref{e:ZK} is invariant under each of these transformations.
Moreover, each of these transformations induces a corresponding transformation for the dependent variables $r_1,\dots,r_3,q$, namely:
\bse
\bea
r_j(\mathbf{x},t) &\mapsto r_j(\mathbf{x}-\mathbf{x}_0,t-t_0), \quad &q(\mathbf{x},t)\mapsto q(\mathbf{x}-\mathbf{x}_0,t-t_0),\\
r_j(\mathbf{x},t) &\mapsto a^2 r_j(a\mathbf{x},a^3 t), \quad &q(\mathbf{x},t)\mapsto q(a\mathbf{x},a^3 t) 
\eea
\ese
for $j=1,2,3$. 
It is straightforward to verify that all these transformations also leave the ZKWS~\eqref{e:ZKWS} invariant. 
For brevity we omit the details.

\paragraph{KdV reduction.}
It is straightforward to see that, when $q=0$ and all quantities are independent of $y$, the ZK-Whitham system reduces
to the Whitham modulation system for the KdV equation, namely
\be
r_{j,t} + V_j\,r_{j,x} = 0\,,\qquad j = 1,2,3\,,
\ee
where $V_1,\dots,V_3$ are the characteristic velocities of the KdV-Whitham system, as above.

\paragraph{Harmonic limit.}
The ZKWS system admits a self-consistent reduction in the harmonic limit $m\to0$ (i.e., $r_2\to r_1$).
In this case, the PDEs for $r_1$ and $r_2$ coincide, and we obtain the reduced $3\times3$ system
\be
\@w_t + A_o\,\@w_x + B_o\,\@w_y = \@0,
\label{e:ZKWS_m=0}
\ee
for the three-component dependent variable $\@w(x,y,t) = (r_1,r_3,q)$, with
\bse
\bea
A_o = \txtfrac13 \left(\begin{array}{ccc}
  3(2 r_1-r_3) + q^2 (4 r_1-r_3) & 0 & 2q (4 r_1^2-2 r_3 r_1+r_3^2) \\
  0 & 3(1+q^2)r_3 & 6 q r_3^2 \\
  - q(1+q^2) & - q(1+q^2) & (1-q^2)(2 r_1+r_3)
\end{array}\right),
\\
B_o = \txtfrac13\left(\begin{array}{ccc}
 2q(r_1-r_3) & 0 & 0 \\
 0 & 0 & 0 \\
 1+q^2 & 1+q^2 & 2q(2 r_1+r_3) 
\end{array}\right).
\eea
\ese

\paragraph{Soliton limit.}
The ZKWS system also admits a self-consistent reduction in the soliton limit $m\to1$ (i.e., $r_2\to r_3$).
The calculations are a bit trickier in this case,
since the entries in the second and third columns of $A$ and $B$ diverge.
As we show next, however, this is not an issue.

Recalling~\eqref{e:mdef}, let $\~m = 1-m = (r_3-r_2)/(r_3-r_1)$,
and write $r_2 = r_3 - \~m\,(r_3-r_1)$.
The limit $m\to1$ corresponds to $\~m\to0$ together with $\~m_x$, $\~m_y$ and $\~m_t$.
We then look at the second and third columns of $A$ and $B$ multiplied by $(r_2,r_3)$.
For the former we have
$a_{i,2}\,r_{2,x} + a_{i,3}\,r_{3,x} = (a_{i,2}+a_{i,3})\,r_{3,x} + a_{i,2}((r_3-r_1)\,\~m)_x$, for $i = 1,\dots,4$. 
with a similar expression for the $y$ derivatives.
Since the singular parts of $a_{i,2}$ and $a_{i,3}$ are exactly equal and opposite,
it is straightforward to verify that one obtains a finite expression in the limit as $\~m\to0$.
The result is the soliton modulation system
\be
\@w_t + A_1\,\@w_x + B_1\,\@w_y = \@0,
\label{e:ZKWS_m=1}
\ee
for the same dependent variables $\@w = (r_1,r_3,q)$ as above, but where the coefficient matrices are now
\bse
\bea
A_1 = \left(\begin{array}{ccc}
 (1+q^2) r_1 & 0 & 2 q r_1^2 \\
 \frac8{15} q^2 (r_1-r_3) & \frac{1}{15}(5(r_1+2r_3) + 3q^2(6r_3-r_1)) 
    & 2q \frac{3r_1^2+48r_3^2 - 6r_1r_3 + q^2(19r_1^2 - 38r_1r_3 + 64r_3^2)}{45(1+q^2)} \\
 -\frac13 q (1+q^2) & -\frac23 q(1 + q^2) & \frac13 (1 - q^2) (r_1+2 r_3) \\
\end{array}\right),
\nonumber\\
\\
B_1 = \left(\begin{array}{ccc}
 0 & 0 & 0 \\
 -\frac{8}{15} q (r_1-r_3) & \frac{8}{15} q (r_1-r_3) & -\frac{8 (5 1+q^2) (r_1-r_3){}^2}{45 (1+q^2)} \\
 \frac{1}{3} (1+q^2) & \frac{2}{3} (1+q^2) & \frac{2}{3} q (r_1+2 r_3) \\
\end{array}\right).
\eea
\ese

\section{Transverse instability of the periodic traveling wave solutions of the ZK equation}
\label{s:stability}

We now show how the ZK-Whitham modulation system derived in section~\ref{s:derivation}
can be applied to study the stability of the periodic traveling wave solutions of the ZK equation
for all $0\le m\le 1$.
We will then compare the predictions of Whitham theory with a numerical evaluation of the instability growth rate,
as well as with an explicit, analytical calculation of the growth rate in the soliton limit.

\subsection{Stability analysis via Whitham theory}
\label{s:Whithamstability}

Recall that, when $r_1,r_2,r_3$ and $q$ are independent of $x,y$ and $t$,
\eqref{e:periodicsolution} is an exact periodic traveling wave solution of \eqref{e:ZK}.
In order to study the stability of such solutions, 
we therefore look for solutions of the ZK-Whitham system \eqref{e:ZKWS} in the form of 
a constant solution $\@r\O0=(r_1\O0,r_2\O0,r_3\O0,q\O0)$ 
plus a small perturbation, namely,
\be
r_j(x,y,t) = r_j\O0 + \delta r_j\O1(x,y,t),\quad j=1,2,3,
\qquad 
q(x,y,t) = q\O0 + \delta q\O1(x,y,t)\,,
\ee
with $0<\delta\ll1$.
Substituting this ansatz into \eqref{e:ZKWS} and neglecting terms $O(\delta^2)$ and smaller,
we then obtain the linearized ZK-Whitham system
\be
\@r\O1_t + A\O0 \@r\O1_x + B\O0 \@r\O1_y = 0,
\label{e:LZKWS}
\ee
since $\@r\O0$ in constant with respect to $x$, $y$ and $t$. 
Here $\@r\O1 = (r_1\O1,r_2\O1,r_3\O1,q\O1)$, while 
$A\O0$ and $B\O0$ denote the $4\times 4$ matrices $A$ and $B$ above evaluated at $\@r=\@r\O0$. 
Since \eqref{e:LZKWS} is a linear system of PDEs with constant coefficients,
it is sufficient to study its plane wave solutions.
We therefore look for solutions in the form
\be
\@r\O1(x,y,t) = \@R\,\e^{i(Kx + Ly - Wt)},
\ee
where $\@R$ is a constant vector,
and $K$, $L$ and $W$ are respectively the perturbation wavenumbers in the $x$ and $y$ directions
and the perturbation's angular frequency.
Substituting this expression into \eqref{e:LZKWS},
the problem above is then transformed into the homogeneous linear system of equations 
$(- W\,I_4 + K\,A\O0 + L\,B\O0)\,\@R = \@0$,
which is equivalent to the eigenvalue problem
\be
(K\,A\O0 + L\,B\O0)\,\@R = W\@R\,.
\label{e:matrixeigenvalueproblem}
\ee
Here $I_4$ is the $4\times4$ identity matrix.  
The eigenvalues (corresponding to nontrivial solutions for $\@R$) 
are the roots of the characteristic polynomial $p(K,L,W) = \det(K\,A\O0 + L\,B\O0 - W\,I_4)$.
In turn, the condition $p(K,L,W)=0$ determines the linearized dispersion relation
$W = W(K,L)$.
If $W\in\Real$, then the constant solution $\@r\O0$ is linearly stable, 
otherwise it is unstable.

By virtue of the scaling invariance of the ZKWS~\eqref{e:ZKWS}, we can set $r_1=0$ and $r_3=1$ without loss of generality,
in which case we simply have $r_2=m$.
Still, for general values of $q$, $K$ and $L$, finding the linearized dispersion $W(K,L)$ involves computing the roots a highly complicated quartic polynomial.
On the other hand, a particularly simple scenario is obtained when $q=0$ (vertical cnoidal waves) and $K=0$ (purely transversal perturbations).
In this case, we simply have $(W/L)^2 = f(m)$, with 
\bea
f(m) = 
\frac4{135}\frac{\big(2(1-m(1-m))\E - (1-m)(2-m)\K\big)\big((1-m)\K^2 - 2 (2-m)\E\K + 3\E^2 \big)}{\E (\K-\E) (\E - (1-m)\K)}\,.
\nonumber\\
\eea

\begin{figure}[t!]
\centerline{\includegraphics[width=0.45\textwidth]{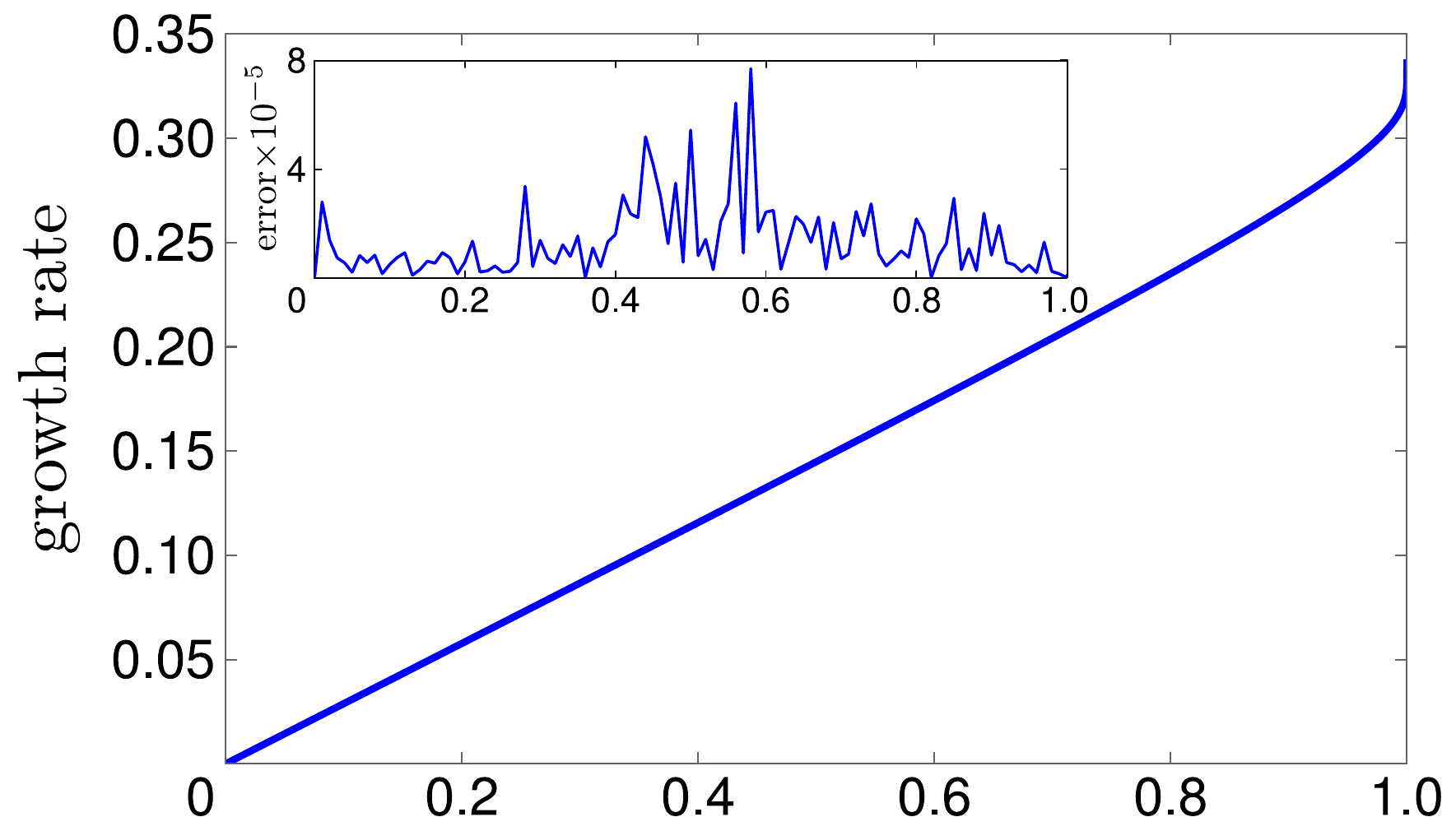}}
\caption{The growth rate of the most unstable transverse perturbation as a function of~$m$ (see text for details).} 
\label{f:2}
\end{figure}

It is straightforward to see that $f(m)<0$ for all $0<m<1$.
Therefore, periodic traveling wave solutions of the ZK equation are linearly unstable with respect to transverse perturbations.
More precisely, 
the above calculations yield the growth rate of the most unstable perturbation as $g(m) = \sqrt{-f(m)}$.
The behavior of $g(m)$ as a function of $m$ is shown in Fig.~\ref{f:2}.
Note that $g(0)=0$ (indicating that the constant solutions are linearly stable), 
and $g(m)$ increases monotonically in $m$,
limiting to the value $g(1) = 4/(3\sqrt{15})\simeq 0.344265$, 
which is the growth rate of unstable perturbations of the soliton solutions of the ZK equation.

The above prediction that the periodic solutions are unstable is consistent with the results of~\cite{Johnson2010}, 
but now we have a fully explicit expression for the instability growth rate,
similar to \cite{ABR2018,Biondini_RSPA2017,Biondini_PRE2017}.
As we show next,
these predictions are in excellent agreement with a numerical calculation of the growth rate (in section \ref{s:linearstability})
as well as with a direct perturbation theory for the soliton solutions (in section~\ref{s:solitonstability}).

\subsection{Stability analysis via linearization of the ZK equation and Floquet-Hill's method}
\label{s:linearstability}

We can validate the predictions of Whitham theory
by studying numerically the linear stability of the periodic solutions of the ZK equation~\eqref{e:ZK} 
and comparing the findings with those obtained via Whitham theory in section~\ref{s:Whithamstability}.

In this case, by analogy with section~\ref{s:Whithamstability}, we look for solutions of the ZK equation~\eqref{e:ZK} 
in two spatial dimensions in the form 
$u(x,y,t) = u_o(x,y,t) + \delta v(x,y,t)$, where $0<\delta\ll1$, and 
where $u_o(x,y,t)$ is an exact periodic traveling wave solution, namely
\bea
u_o(x,y,t) = (1+q^2)((r_1 - r_2 + r_3) + 2(r_2-r_1)\cn^2(2K_m Z,m)) 
\nonumber\\
\kern6em
 = (1+q^2)\,\big[r_1 - r_2 + r_3 + 2(r_2-r_1)\cn^2\big({\sqrt{(r_3-r_1)/6}(x+qy-Vt)}/\epsilon\big)\big]\,,
\label{e:cnoidal_perturbationtheory}
\eea
$Z = k(x + qy - Vt)/\epsilon$ is the fast variable defined in~\eqref{e:Zfast}, 
$V = \omega/k$ and $k$ and $\omega$ are as in~\eqref{e:klomega}.
Substituting this ansatz in~\eqref{e:ZK}, to leading order in $\delta$
we obtain a linearized ZK equation:
\be
v_t + (u_0 v)_x + \epsilon^2 (\Delta v)_x = 0\,.
\label{v-pert-eqn}
\ee
To obtain the correct balance of terms in $\epsilon$ and $\delta$, 
we use the following ansatz for $v(x,y,t)$:
\be
v(x,y,t) = w(2K_m Z)e^{(i\zeta y + \lambda t)/\epsilon}
\ee
which implies
\bse
\bea
v_x = \frac{2K_m k}{\epsilon} v_Z, \quad 
v_y = \frac{2K_m l}{\epsilon} v_Z + \frac{i\zeta}{\epsilon}v, \quad 
v_t = -\frac{2K_m \omega}{\epsilon} v_Z + \frac{\lambda}{\epsilon}v,
\\
v_{yy} = \frac{4K_m^2 l^2}{\epsilon^2}v_{ZZ} + \frac{4iK_m \zeta l}{\epsilon^2}v_Z 
	- \frac{\zeta^2}{\epsilon^2}v + O\left(\frac{1}{\epsilon}\right), \quad
v_{xx} = \frac{4K_m^2 k^2}{\epsilon^2} v_{ZZ},
\eea
\ese
with $l = qk$ as before.
Then, to leading order in $\epsilon$, \eqref{v-pert-eqn}~yields
\be
-2K_m \omega v_Z + \lambda v + 2K_m k(u_o v)_Z + 8K_m^3 k^3(1+q^2)v_{ZZZ} + 8iK_m^2 \zeta k l v_{ZZ} - 2K_m k\zeta^2  v_Z = 0\,,
\label{e:perturbedZK0}
\ee
which can be written as the linear eigenvalue problem
\bse
\bea
\mathcal{L}_o v = \lambda v\,,
\label{e:lineareigenvalueproblem0}
\\
\noalign{\noindent with}
\mathcal{L}_o = 2K_m \omega\partial_Z - 2K_m k\, \partial_Zu_0 - 8K_m^3 k^3(1+q^2)\partial_Z^3
	- 8iK_m^2 \zeta k^2q \partial_Z^2 + k \zeta^2\partial_Z\,. 
\eea
\ese
Explicitly, using the definition of $k$, \eqref{e:perturbedZK0} is
\bea
-\sqrt{r_3-r_1} V v_Z + \~\lambda v + \sqrt{r_3-r_1}(u_o v)_Z + (r_3-r_1)^{3/2}[(1+q^2)/6]\,v_{ZZZ}
\nonumber\\
\kern16em 
  + 2iq\,[(r_3-r_1)/\sqrt6]\,\zeta v_{ZZ} - \sqrt{r_3-r_1}\,\zeta^2 v_Z = 0\,,
\label{e:perturbedZK1} 
\eea
where $\~\lambda = \sqrt6\,\lambda$.
To compare the results of this perturbation expansion with the predictions of Whitham theory, 
we set $r_1=0$ and $r_3=1$, implying $r_2 = m$, 
and we take $q=0$. 
Then \eqref{e:perturbedZK1} yields
\be
-V v_Z + \~\lambda v + (u_o v)_Z + (1/6)v_{ZZZ} - \zeta^2 v_Z = 0 \,.
\ee
Equivalently, the eigenvalue problem \eqref{e:lineareigenvalueproblem} becomes 
\bse
\bea
\mathcal{L} v = \~\lambda v\,,
\label{e:lineareigenvalueproblem}
\\
\noalign{\noindent where}
\mathcal{L} = V\,\partial_Z - \partial_Z \,u_o - (1/6)\partial_Z^3 + \zeta^2\partial_Z\,.
\eea
\ese
We compute the eigenvalues $\~\lambda$ of $\mathcal{L}$ numerically for each $0\le m<1$ using Floquet-Hill's method \cite{deconinckkutz}. 
The difference between the resulting values and those obtained via Whitham theory shown in the inset of Fig.~\ref{f:2},
which demonstrates excellent agreement between the two approaches.
(Note that the numerical values of the discrepancy between the two approaches depend somewhat 
on the value of $\zeta$ chosen, since the latter affects the accuracy of 
the numerical scheme.  The values in Fig.~\ref{f:2} were obtained with $\zeta = 0.0005$.)
Note however that, 
unlike the present approach,
Whitham theory yields an analytical expression for the instability growth rate.

\subsection{Analytical stability theory for soliton solutions}
\label{s:solitonstability}

As a final test for the predictions of Whitham theory, we now calculate the instability growth rate for the 
soliton solutions analytically.
That is, we look for perturbed solution in the following form:
\be
u(x,y,t) = u_c(\xi) + U(\xi)e^{i\zeta y + \lambda t},
\label{e:solitonperturbation}
\ee
where $u_c(\xi)$ is the solitary wave solution [i.e., the limit $m\to 1$ of \eqref{e:cnoidal_perturbationtheory}], 
and the second term in~\eqref{e:solitonperturbation} describes purely transversal perturbations.
For concreteness, we choose $r_1=0$ and $r_2=r_3=6c$
(with the specific parametrization chosen so as to simplify the calculations that follow,
similarly to \cite{Pelinovsky2018}), 
and $q=0$. 
We then have $2K_m Z = \sqrt{c}(x + qy - 4ct) = \sqrt{c}\xi$, 
where $\xi = x-4ct$,
and, as per \eqref{e:solitonsolution}, 
\be
u_c(\xi) = 12 c \sech^2(\sqrt{c}\xi)\,.
\ee
We write the ZK equation \eqref{e:ZK} in the soliton comoving reference frame $(\xi,y,t)$, 
which reduces the problem to the analysis of ordinary differential equations.
We then look for a formal asymptotic expansions in $\zeta$ for $\lambda$ and $U$ near $\zeta=0$, namely:
\bse
\bea
\lambda = \lambda_1 \zeta + \lambda_2 \zeta^2 + O(\zeta^3),
\\
U(\xi) = U_0(\xi) + \lambda_1 \zeta U_1(\xi) + \lambda_2 \zeta^2 U_1(\xi) + \zeta^2 U_2(\xi) + O(\zeta^3).
\eea
\ese
We should point out the similarities and the differences between the present approach and that of \cite{Pelinovsky2018}.
The perturbation expansion above is similar in spirit to that in \cite{Pelinovsky2018}. 
However, \cite{Pelinovsky2018} studied the stability of solitary waves with speed close to the critical speed of propagation,
whereas in this case we are studying the stability near zero transverse wavenumbers.

Substituting this ansatz into the ZK equation written in the comoving reference frame, 
at leading order we obviously simply recover an ordinary differential equation that yields the soliton solution:
\be
u_c'' + \txtfrac12 u_c^2 - 4cu_c = 0\,,
\ee
where primes denote differentiation with respect to $\xi$.
Then the eigenvalue problem for $\lambda$ can be written as
\be
\partial_\xi (M + \zeta^2)U = \lambda U, 
\label{e:lambda}
\ee
where
\be
M = -\partial_\xi^2 + 4c - 12c\sech^2(\sqrt{c}\xi).
\ee
We can write
\be
\lambda U = \zeta(\lambda_1 U_0) + \zeta^2 (\lambda_2 U_0 + \lambda_1^2 U_1) + O(\zeta^3)
\ee
and
\be
\partial_\xi MU = \partial_\xi MU_0 + \lambda_1 \zeta \partial_\xi MU_1 
+ \lambda_2 \zeta^2 \partial_\xi MU_1 + \zeta^2 \partial_\xi MU_2 + O(\zeta^3).
\ee
At $O(1)$ in $\zeta$ of the eigenvalue problem~\eqref{e:lambda} we have
\be
\partial_\xi (MU_0) = 0,
\ee
which yields $U_0 = u_c'(\xi)$. 
At $O(\zeta)$ we have
\be
\partial_\xi MU_1 = U_0\,,
\ee
i.e.,
\be
\big[-\partial_\xi^2 +4c-12c\sech^2(\sqrt{c}\xi)\big]\,U_1 = 12c\sech^2(\sqrt{c}\xi)\,.
\ee
It is straightforward to see that the above ODE admits the solution
\be
U_1(\xi) = \frac{3}{4}\sech^2(\sqrt{c}\xi)\left(-4+(5+4\sqrt{c}\xi)\tanh(\sqrt{c}\xi)\right)\,.
\ee
Then, and finally, at $O(\zeta^2)$ we have
\be
\partial_\xi MU_2 + \lambda_2 U_0 + \partial_\xi U_0 = \lambda_2 U_0 + \lambda_1^2 U_1,
\ee
or equivalently
\be
\partial_\xi MU_2 = \lambda_1^2 U_1 - \partial_\xi^2 u_c \,.
\label{e:O2pert}
\ee
The Fredholm solvability condition requires the right-hand side of~\eqref{e:O2pert} to be orthogonal 
to the kernel of the adjoint of the operator in the left-hand side in order for~\eqref{e:O2pert} to admit solutions.
Since $M$ is self-adjoint, the adjoint of $\partial_\xi M$ is simply $M\partial_\xi$.
The kernel in question is thus spanned by $u_c$. 
Therefore the resulting constraint is
\be
\lambda_1^2 \int_\Real u_1 u_c\,\d\xi = \int_\Real u_c u_c''\,\d\xi\,,
\ee
and this condition determines~$\lambda_1$.
The integrals in the above conditions are given by, respectively,
\be
\int_\Real u_1 u_c\,\d\xi = -36\sqrt{c}, \quad 
\int_\Real u_c u_c''\,\d\xi = - \int (u_c')^2 d\xi = -\frac{768}{5}c^{5/2}.
\ee
Their ratio then gives $\lambda_1$ as
\be
\lambda_1 = \frac{8}{\sqrt{15}}c,
\label{e:lambda1analytical}
\ee
In order to compare this result with Whitham theory, 
note that in that case we took $r_3=1$, implying $c = 1/6$,
which then yields $\lambda_1 = 4/(3\sqrt{15})$,
which is in perfect agreement with the results of section~\ref{s:Whithamstability}.

We should note that the above formalism can be generalized in a relatively straightforward way to compute the instability growth rate for all periodic solutions of the ZK equation.  
However, the corresponding calculations are somewhat more involved,
and at the moment they have not yet led to a closed-form result similar to~\eqref{e:lambda1analytical}.
For brevity, they are therefore deferred to a future publication.

\section{Concluding remarks}
\label{s:conclusions}

In summary, we have derived the ZK-Whitham system (ZKWS), i.e., the system of Whitham modulation equations 
for the periodic solutions of the Zakharov-Kuznetsov equation.
The ZKWS shares some similarities with the KP-Whitham system, i.e., the system of modulation equations for the KP equation.
Both are first-order systems of PDEs of hydrodynamic type,
and both systems involve three time evolution equations for the Riemann-type variables $r_1,\dots,r_3$ 
plus a fourth time evolution equation for the local slope parameter $q = k_2/k_1$.
At the same time, there are some important differences between the two modulation systems.
Most importantly, the fact that the ZKWS
comprises only four PDEs, whereas the KP-Whitham system contains an additional PDE (which does not contain time derivatives) 
for an auxiliary field.  
(As mentioned in \cite{Biondini_RSPA2017}, the presence this fifth PDE is essential for the system to correctly capture
the dynamics of solutions of the KP equation.)
We also studied the harmonic and soliton limits of the ZKWS, and we used the ZKWS to study the transverse stability of the periodic traveling wave solutions, showing that all such solutions are unstable to transverse perturbations.

The instability of such solutions raises the interesting question of whether the ZK equation admits any exact solutions
describing stable two-dimensional wave patterns.
Another interesting question is whether the ZKWS can be used to study time evolution problems
similarly to what was done in \cite{NLTY21,PRSA22,JFM21}
for the KP equation.
The situation for the ZK equation is different because its periodic solutions are unstable.
Still, it is well known that Whitham modulation equations can be very useful even when the underlying solutions of the PDE 
are unstable and the system is not hyperbolic (e.g., as in the case of the modulational instability of constant solutions of the focusing one-dimensional NLS equation \cite{PRL116p043902,elgurevich,kamchatnov}).
A natural question is therefore where special solutions of the ZKWS could be useful to capture certain features of the time evolution of solutions of the ZK equation.

Obviously it would also be interesting to study the ZKWS as a (2+1)-dimensional hydrodynamic system on its own, 
independently of its connection with the ZK~equation.
On that note, we point out that, similarly to what happens with the KP equation \cite{arXiv2303.06436}, 
solutions of the ZKWS describe the modulation of solutions of the ZK equation only when the initial conditions for the 
ZKWS are consistent with the third conservation of waves equation, i.e., the constraint $k_y = (qk)_x$.
As with the KP equation \cite{Biondini_RSPA2017}, 
it is straightforward to show that if this condition is satisfied at time zero,
the ZKWS ensures that it is preserved by the time evolution.
A related question concerns the possible integrability of the ZKWS.
Since the ZK equation is not integrable, one would not expect the ZKWS to be integrable.
Nonetheless, it is possible that certain reductions such as the harmonic limit and the soliton limit, 
could nonetheless be integrable.

All of these questions are left for future investigation, and 
it is hoped that the results of this work and the above remarks will stimulate further study on these topics.

\section*{Acknowledgments}

We are indebted to Dmitry Pelinovsky for his help in calculating the soliton instability growth rate 
in section~\ref{s:solitonstability}.
We also thank Gigliola Staffilani for many interesting conversations.
This work was partially supported by the National Science Foundation under grant number DMS-209487.

\def\title#1{``#1''}
\let\reftitle=\title

\end{document}